\documentclass[conference]{IEEEtran}
\usepackage{float}
\IEEEoverridecommandlockouts
\usepackage[utf8]{inputenc}
\UseRawInputEncoding 
\usepackage{cite}
\usepackage{amsmath,amssymb,amsfonts}
\usepackage{amsthm}
\usepackage{amsthm}
\theoremstyle{remark}

\usepackage{algorithm}
\usepackage{algorithmic}
\usepackage{graphicx}
\usepackage{textcomp}
\usepackage{xcolor}
\usepackage{stfloats}
\usepackage{booktabs}
\def\BibTeX{{\rm B\kern-.05em{\sc i\kern-.025em b}\kern-.08em
    T\kern-.1667em\lower.7ex\hbox{E}\kern-.125emX}}
\begin{document}

\title {Transfer to Sky: Unveil Low-Altitude Route-Level Radio Maps via Ground Crowdsourced Data\\

}
\author{
\IEEEauthorblockN{
Wenlihan Lu\IEEEauthorrefmark{1}\IEEEauthorrefmark{2},
Huacong Chen\IEEEauthorrefmark{2},
Ruiyang Duan\IEEEauthorrefmark{2},
Weijie Yuan\IEEEauthorrefmark{3},
Shijian Gao\IEEEauthorrefmark{1}
}
\IEEEauthorblockA{
\IEEEauthorrefmark{1} Hong Kong University of Science and Technology (Guangzhou), Guangzhou, China \\
\IEEEauthorrefmark{2} Meituan, Shenzhen, China 
\IEEEauthorrefmark{3} Southern University of Science and Technology, Shenzhen, China \\
}

}

\maketitle

\begin{abstract}
The expansion of the low-altitude economy is contingent on reliable cellular connectivity for unmanned aerial vehicles (UAVs). A key challenge in pre-flight planning is predicting communication link quality along proposed and pre-defined routes, a task hampered by sparse measurements that render existing radio map methods ineffective. This paper introduces a transfer learning framework for high-fidelity route-level radio map prediction. Our key insight is to leverage abundant crowdsourced ground signals as auxiliary supervision. To bridge the significant domain gap between ground and aerial data and address spatial sparsity, our framework learns general propagation priors from simulation, performs adversarial alignment of the feature spaces, and is fine-tuned on limited real UAV measurements. Extensive experiments on a real-world dataset from Meituan show that our method achieves over 50\% higher accuracy in predicting Route RSRP compared to state-of-the-art baselines.
\end{abstract}


\section{Introduction}
Unmanned Aerial Vehicles (UAVs) are becoming critical infrastructure for the emerging low-altitude economy~\cite{monitoring}. Reliable communication links are essential for UAVs to execute functions like video transmission, state reporting, and emergency control~\cite{erergency}. However, urban obstacles and non-line-of-sight (NLOS) conditions often cause severe connectivity degradation. To mitigate this, utilizing route-level radio maps during the pre-flight planning stage is a critical step. By evaluating spatial variations in received signal power, these radio maps guide the selection of optimal flight paths, thereby ensuring both communication reliability and operational safety~\cite{radiomap}.

\begin{figure}[!t]
    \centering
\includegraphics[width=1\linewidth]{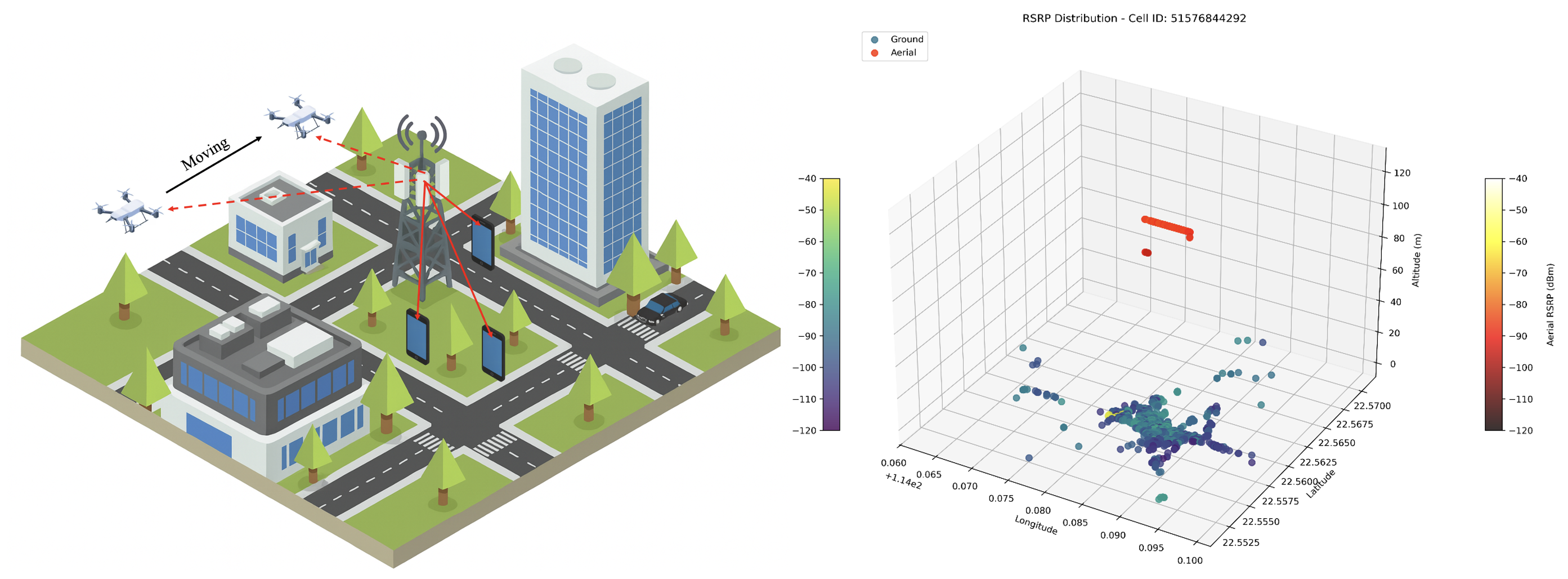}
    \caption{Illustration of a single-cell scenario with UAV and ground users. The right figure shows the 3D distribution of RSRP for aerial and ground measurements.}
    \label{fig:scene}
\end{figure}

While radio map reconstruction has received significant attention in recent years, the specific problem of constructing \textit{route-level} radio maps remains largely unexplored and proves to be particularly challenging. Existing methods, including interpolation~\cite{matrixInterpolation}, kriging~\cite{kriging}, and learning-based models~\cite{machineradio, sip2net}, typically rely on two key assumptions: dense, uniformly distributed measurements~\cite{rmgen} and accurate environmental priors~\cite{radioUnet}. However, these assumptions are invalidated by the realities of UAV operation. UAVs are constrained to predefined flight routes, and dynamic cell selection mechanisms~\cite{cellswitch} result in sparse, irregular, and discontinuous aerial measurements, as illustrated in Fig.~\ref{fig:scene}. This unique sparsity pattern fundamentally undermines the foundations of traditional reconstruction techniques, rendering them ineffective for this critical application.

To mitigate data scarcity, some studies attempt to equip UAVs with network scanners for active aerial surveys. These sparse measurements are typically collected by UAVs as they fly along trajectories planned by traditional~\cite{spectrumsurveying} or reinforcement learning-based algorithms~\cite{activeradiomap}. While these methods improve coverage estimation, their high energy cost, limited flight endurance, and airspace constraints hinder scalability for city-level deployments. Given the inability of existing methods to cope with sparse aerial data, this paper proposes to leverage abundant and easily attainable ground crowdsourced measurements to infer aerial coverage, as both share similar propagation environments. Nevertheless, it is worth noting that a significant domain gap exists: ground samples are densely clustered near base stations and are predominantly NLoS, whereas aerial ones are sparse, scattered, and exhibit a higher probability of Line-of-Sight (LoS) links. as shown in Fig.~\ref{fig:scene}. This discrepancy prevents reliable direct mapping from ground to air, especially under limited UAV observations.

This work presents the first study on recovering route-level low-altitude radio maps from abundant ground crowdsourced measurements, enabling proactive assessment of UAV cellular connectivity along planned logistics routes. To overcome the dual challenges of aerial data scarcity and the risk of overfitting on sparse UAV samples, we propose a three-stage transfer learning framework. Our approach first leverages large-scale simulation-generated data to pretrain a dual-transmitter U-Net for masked radio map recovery. This self-supervised task builds robust propagation priors and spatial awareness of urban morphology. Subsequently, an adversarial domain adaptation stage aligns the feature distributions between simulated and real-world data, effectively narrowing the sim-to-real gap. Finally, a lightweight decoder finetuning is performed using sparse UAV measurements to calibrate the model toward real propagation characteristics. Experiments on Meituan\textquotesingle s operational UAV datasets demonstrate a 51.9\% improvement in route-level radio map RSRP prediction accuracy over state-of-the-art baselines.

\section{Problem Statement}

This work aims to predict the route-level RSRP for planned UAV trajectories under a specific cell ID by leveraging abundant ground measurements to infer aerial RSRP distributions. Although UAV trajectories may cross multiple cells, we focus on a critical first step toward a more general solution. We assume temporal consistency between the ground and aerial datasets, as large-scale fading is mainly determined by static environmental factors~\cite{temporal}.

For each cell, we define two measurement sets: a ground set and an aerial set. 
The former is denoted as $\mathcal{D}_g$, samples collected from user equipment (UE) operating on the ground near the base station. 
Each element in $\mathcal{D}_g$ is given by $(\mathbf{x}_{g,n}, p_{g,n})$, 
where $\mathbf{x}_{g,n} = (x_n, y_n, z_n)$ represents the three-dimensional location of the $n$-th ground sample (\textit{with altitude $z_n$ typically below 50~m}), 
and $p_{g,n}$ represents the RSRP. 
These elements are crowdsourced from mobile devices that periodically report their GPS coordinates, serving cell identifiers, and RSRP values to the network. 
Consequently, $\mathcal{D}_g$ is usually dense but spatially non-uniform, because most samples are concentrated near the base stations.

Similarly, the aerial measurement set, $\mathcal{D}_a$, comprises UAV sampling records collected at higher altitudes. 
Each element $(\mathbf{x}_{a,m}, p_{a,m})$ includes the UAV\textquotesingle s 3D position $\mathbf{x}_{a,j} = (x_m, y_m, z_m)$ and the corresponding RSRP value $p_{a,m}$. 
Due to the high operational cost and limited accessibility of UAV flights, $\mathcal{D}_a$ is extremely sparse compared with $\mathcal{D}_g$, 
typically satisfying $|\mathcal{D}_a| / |\mathcal{D}_g| < 10^{-3}$.

{
Ideally, we aim to learn a mapping function $f(\mathbf{x}_j; \mathcal{D}_g^c, \theta)$ 
that integrates contextual information from the ground measurement set $\mathcal{D}_g^c$ 
and generalizes to predict the RSRP at UAV route coordinates in 3D space. 
The desired mapping can be formulated as the following reconstruction objective:
\begin{equation}
    \theta^* = \arg\min_{\theta}
    \frac{1}{|\mathcal{D}_a|} \sum_{(\mathbf{x}_{a,m}, p_{a,m}) \in \mathcal{D}_a}
    \left\| f(\mathbf{x}_{a,m}; \mathcal{D}_g, \theta) - p_{a,m} \right\|^2,
    \label{eq:reconstruction_obj}
\end{equation}
Learning a reliable ground-to-air mapping is non-trivial due to the scarcity of aerial data and the intrinsic domain discrepancy caused by differing propagation environments. The core challenge is to leverage rich ground data under limited aerial measurements supervision. 
}

\section{Proposed Framework}
\label{sec:method}

To address the challenges of ground-to-air radio map recovery, we design a three-stage framework, as illustrated in Fig.~\ref{fig:framework}. The framework first pretrains a radio map reconstruction network using simulated augmented data, aims to learn general propagation priors, then bridges the gap between simulated and real world data via adversarial domain adaptation, and finally finetunes the model with real data to adapt to real world propagation characteristics.

\subsection{Pretrain via Synthetic Augment Data}
Given the scarcity of aerial measurements, the first stage leverages ray-tracing simulation to provide abundant training data and establish transferable knowledge of radio propagation. The objective is to endow the model with a foundational understanding of both large scale and fine grained propagation behaviors in urban environments, which can later be adapted to real world conditions.

\begin{figure*}[!t]
    \centering
    \includegraphics[width=1.00\textwidth]{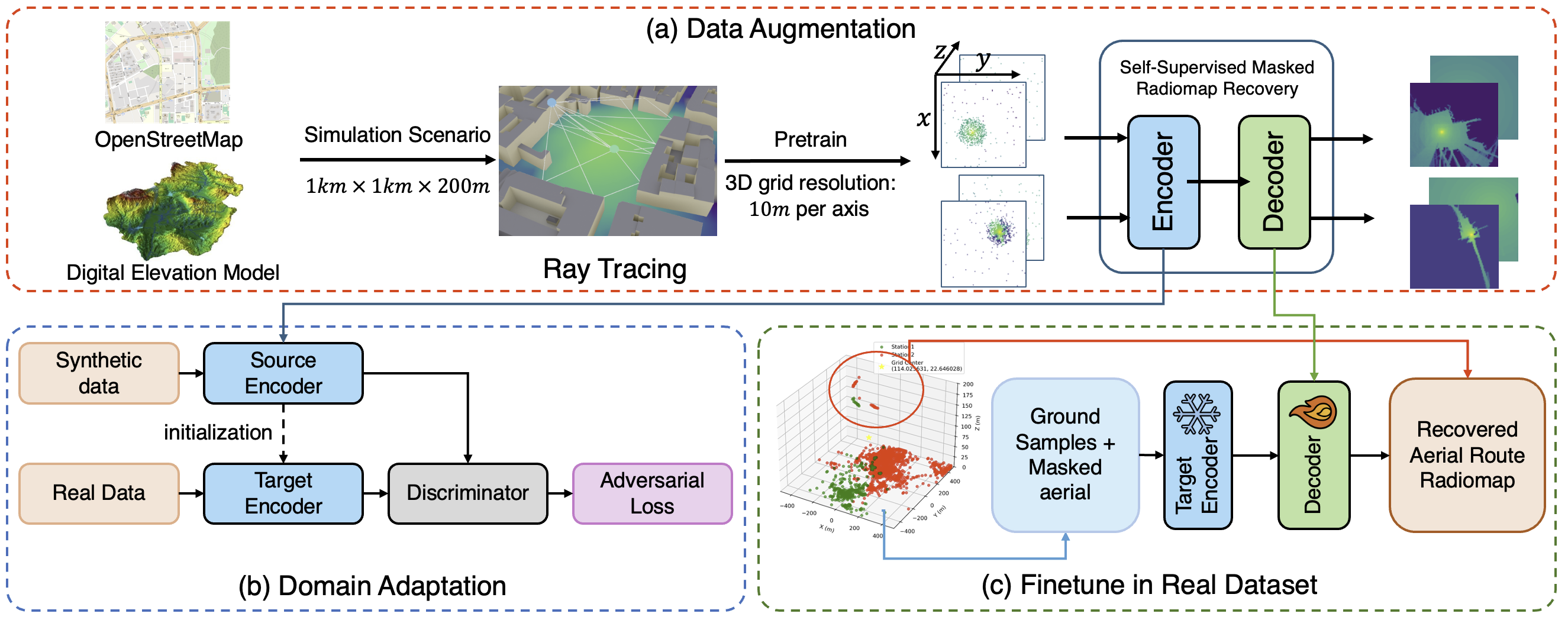}
    \caption{An illustration of the proposed framework.
}
    \label{fig:framework}
\end{figure*}

{As illustrated in Fig.~\ref{fig:framework} (a), the large-scale training data are synthesized by leveraging publicly accessible geospatial resources. Specifically, building footprints and height attributes are extracted from the OpenStreetMap (OSM) database, while terrain elevation information is obtained from digital elevation models (DEM)~\cite{OSM}. The integration of these two layers enables the reconstruction of city scale 3D environments that faithfully represent both urban morphology and topographic variations. These reconstructed environments are subsequently imported into the Sionna ray-tracing engine~\cite{sionna} to simulate large scale 3D radio maps, which encapsulate rich electromagnetic propagation characteristics that are otherwise infeasible to acquire from limited UAV measurements.}

With the generated data $\mathcal{D}_s$, we pretrain a U-Net style encoder-decoder network through a self-supervised task termed \emph{masked radio map recovery}. For each complete 3D radio map $X_s \in \mathcal{D}_s$, we construct a binary mask $M \in \{0,1\}^{H \times W \times D}$ whose entries decide whether the RSRP value at $(h,w,d)$ is retained or removed, yielding the masked input
\begin{equation}
X_s' = M \odot X_s,
\end{equation}
where $\odot$ denotes the Hadamard product. To reflect realistic acquisition, masking is spatially biased rather than uniform. 
Therefore, a sampling rule is tailored per the distance threshold. That is, voxels closer to the nearest transmitter are retained with a higher probability,
while distant voxels are sampled less frequently.
Let $d(\mathbf{v}) = \|\mathbf{v} - tx\|_2$ denote the distance from voxel $\mathbf{v} = (h,w,d)$ to its transmitter. The retention probability is expressed as
\begin{equation}
P(M_{i,j,k}=1) =
\begin{cases}
p_{\text{near}}, & d(\mathbf{v}) \le r_1, \\[3pt]
p_{\text{mid}},  & r_1 < d(\mathbf{v}) \le r_2, \\[3pt]
p_{\text{far}},  & d(\mathbf{v}) > r_2,
\end{cases}
\label{eq:mask_prob_piecewise}
\end{equation}
where $r_1 < r_2$ are distance thresholds, and $0 < p_{\text{far}} < p_{\text{mid}} < p_{\text{near}} < 1$
controls the sampling density across in different distance ranges.
This strategy concentrates samples near base stations and sparsifies distant regions, aligning the pretraining data distribution with real-world measurement patterns.

\textit{Remark on Sampling Probability Design}:
The probabilities $(p_{\text{near}}, p_{\text{mid}}, p_{\text{far}})$ are fixed to mimic real measurement densities in cellular networks.
Voxels near transmitters are sampled more frequently ($p_{\text{near}}$) to reflect dense user distributions,
while $p_{\text{mid}}$ and $p_{\text{far}}$ represent sparser sampling at greater distances.

Given the masked input $X_s'$, the encoder $f_{\text{enc}}(\cdot)$ extracts high-level spatial representations. They will be fed into the decoder $f_{\text{dec}}(\cdot)$ to output a reconstructed radio map $\hat{X}_s$. 
In practice, a single base station occupies a limited spatial coverage, whereas multiple nearby transmitters may share and complement certain regions. 
By aggregating their coverage patterns, the network is expected to learn a more holistic representation of the 3D radio environment, enabling accurate reconstruction across diverse propagation conditions.

\begin{figure}[t]
    \centering
    \includegraphics[width=1\linewidth]{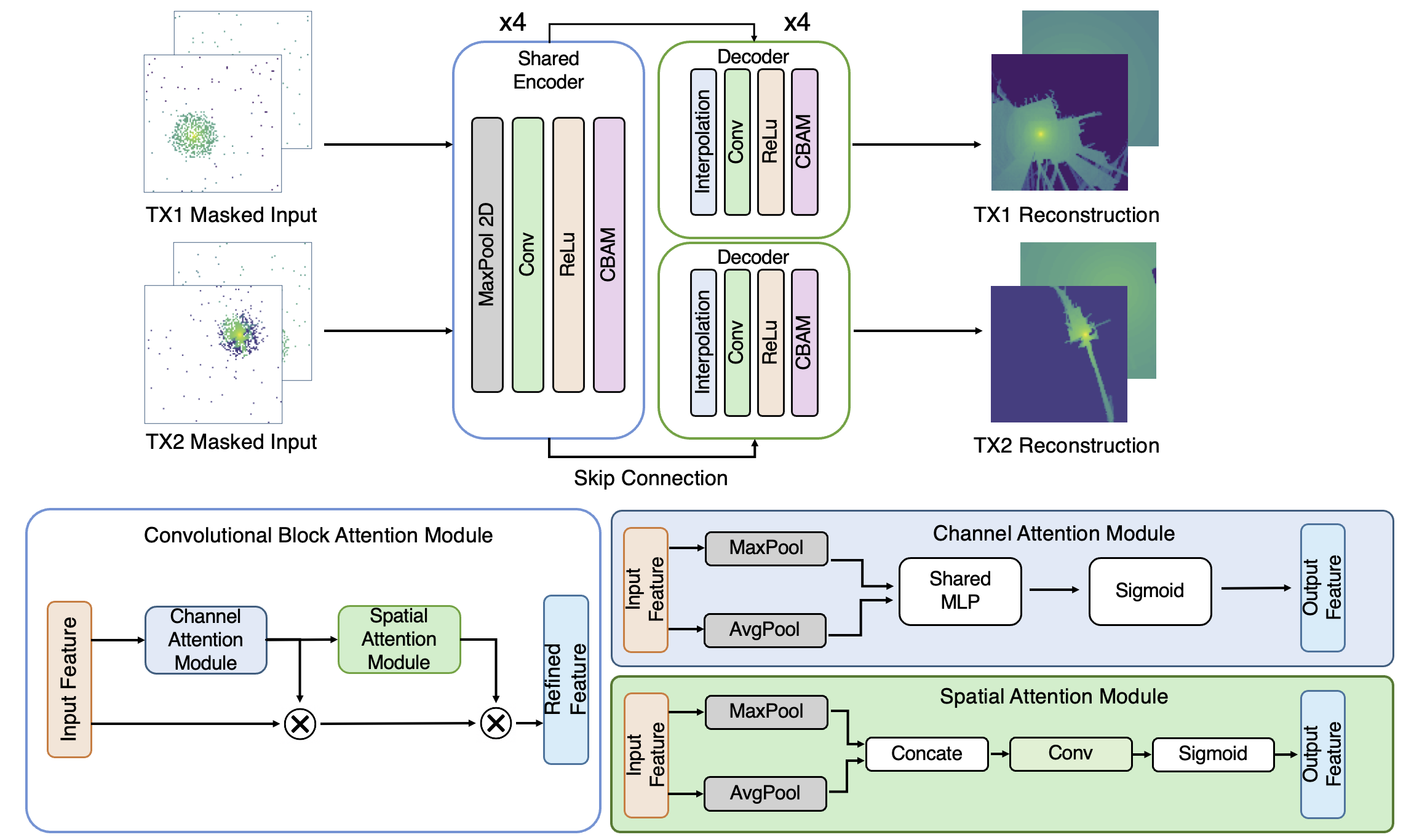}
    \caption{U-Net with a shared encoder and attention-enhanced, transmitter-specific decoders for masked radio map recovery.}
    \label{fig:Net}
\end{figure}

\begin{figure}
    \centering
    \includegraphics[width=1\linewidth]{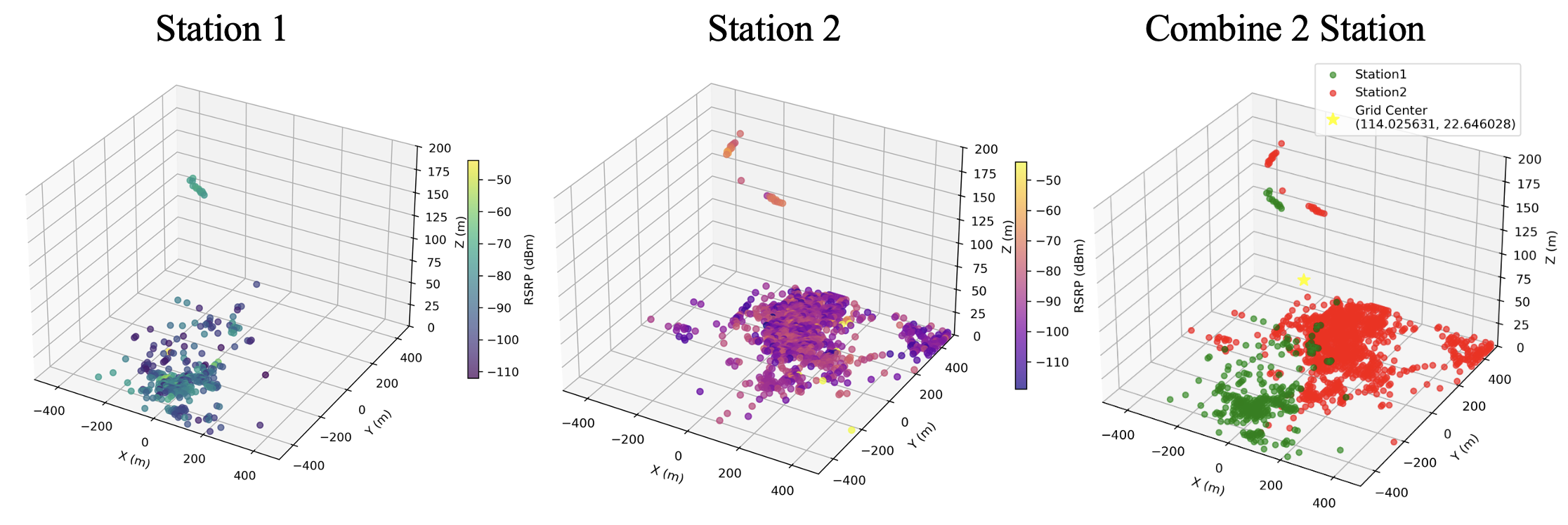}
    \caption{Complementary 3D coverage of two adjacent transmitters and their combined grid representation.}
    \label{fig:2tx}
\end{figure}

As illustrated in Fig.~\ref{fig:2tx}, two adjacent transmitters exhibit overlapping coverage patterns, demonstrating this complementary property. 
To harness this spatial diversity, we design a dual-transmitter U-Net featuring a shared encoder and two transmitter-specific decoders (see Fig.~\ref{fig:Net}), and pretrain it on a self-supervised task of masked radio map recovery. 
The Convolutional Block Attention Module (CBAM)~\cite{cbam} is integrated to adaptively refine feature representations along both the channel and spatial dimensions. 
CBAM enables the network to effectively exploit the complementary spatial characteristics of adjacent transmitters, 
thereby enhancing the fusion of their propagation features in overlapping areas.

For this task, we utilize the simulated dataset $\mathcal{D}_s$. 
For each pair of adjacent transmitters indexed by $(i,j)$, we generate two masked input radio maps 
$X_s^{(i)}$ and $X_s^{(j)}$, where $(i,j)\in\mathcal{E}$ and $\mathcal{E}$ denotes the set of adjacent-transmitter index pairs.
Accordingly, the voxel-retention probability in each radio map follows Eq.~(\ref{eq:mask_prob_piecewise}), 
with distance measured to the corresponding transmitter indexed by $i$ or $j$.
A shared encoder $f_{\mathrm{enc}}$ is followed by two parameter-separated but architecturally identical decoder heads:
\begin{equation}
\begin{aligned}
\widehat{X}_s^{(i)} &= f_{\text{dec}}^{(i)}\!\big(f_{\mathrm{enc}}(X_s^{\prime(i)});\;\boldsymbol{\theta}_{\text{dec},i}\big),\\
\widehat{X}_s^{(j)} &= f_{\text{dec}}^{(j)}\!\big(f_{\mathrm{enc}}(X_s^{\prime(j)});\;\boldsymbol{\theta}_{\text{dec},j}\big).
\end{aligned}
\end{equation}
Two decoder heads $f_{\text{dec}}^{(i)}(\cdot)$ and $f_{\text{dec}}^{(j)}(\cdot)$ then reconstruct the complete 3D radio maps for their respective transmitters.
Here, the reconstruction refers to the simulated full radio map within each transmitter’s coverage region, representing the volumetric RSRP field in the pretraining environment. 
This self-supervised objective is designed to learn transferable propagation priors and spatial dependencies from simulated environments, rather than route-specific signal patterns. 
Accordingly, the pretraining loss is defined as the joint reconstruction error over adjacent transmitter pairs:
\begin{equation}
\mathcal{L}_{\text{pretrain}}
=\big\|\widehat{X}_s^{(i)}-X_s^{(i)}\big\|_2^2
+\big\|\widehat{X}_s^{(j)}-X_s^{(j)}\big\|_2^2.
\label{eq:pretrain}
\end{equation}
This dual-cell design encourages the encoder to extract transferable propagation features that generalize across different transmitters while preserving cell-specific decoding capability through the decoders.

\subsection{Adversarial Domain Adaptation}
With the model pretrained on simulated radio maps (Sec.~III-A) to capture generalizable propagation priors, we address the \textit{sim-to-real} gap by contrasting the two domains. The simulated source domain \(D_s\) contains dense, label-rich, and low-noise radio maps under controlled conditions, while the real-world target domain \(D_t\) comprises sparse, irregular, and noisy UAV measurements with scarce labels. This discrepancy motivates an adaptation strategy to (i) transfer priors from \(D_s\), (ii) align representations without target labels, and (iii) preserve pretrained knowledge.

Motivated by the above requirements and inspired by \cite{adda}, we employ Adversarial Discriminative Domain Adaptation (ADDA) to bridge the distribution gap between simulated and real domains. 
ADDA learns a target encoder \(f_{\text{enc}}^{T}\) by adversarially aligning the latent representations of the real target domain \(D_t\) to those of the simulated source domain \(D_s\) via a domain discriminator, while keeping the source encoder \(f_{\text{enc}}\) fixed. 
In our dual-TX U\mbox{-}Net framework shown in Fig.~\ref{fig:Net}, this alignment is enforced at the shared-encoder bottleneck \(Z\), whereas the two transmitter-specific decoders \(f_{\text{dec}}^{(i)}\) and \(f_{\text{dec}}^{(j)}\) remain cell-aware and frozen. 
Such a design alleviates covariate shift without overwriting the simulation priors, making it suitable for label-scarce target domains and robust against sampling imbalance or feature-distribution discrepancies encountered in real UAV operations.

In practice, the target encoder is initialized from the pretrained shared encoder, 
\(f_{\text{enc}}^{T} \leftarrow f_{\text{enc}}\). 
During adaptation, the discriminator operates on the bottleneck features \(Z\) extracted from the masked 3D grids of both TX streams, and we update only \(f_{\text{enc}}^{T}\) and the discriminator while freezing the source encoder and both TX-specific decoders \(f_{\text{dec}}^{(i)}\) and \(f_{\text{dec}}^{(j)}\). 
The discriminator aims to distinguish the domain origin of each feature, whereas the target encoder learns to deceive it by generating source-like embeddings. 
Their adversarial training is governed by the binary cross-entropy (BCE) objective:
\begin{align}
\mathcal{L}_{D} 
&= - \mathbb{E}_{X_s \sim \mathcal{D}_s}\!\left[\log D(f_{\text{enc}}(X'_s))\right] \nonumber \\
&\hspace{2em} - \mathbb{E}_{X_t \sim \mathcal{D}_t}\!\left[\log (1 - D(f_{\text{enc}}^{T}(X'_t)))\right].
\end{align}
Conversely, the target encoder $f_{\text{enc}}^{T}$ is trained to generate target features that can fool the discriminator by minimizing:
\begin{equation}
\mathcal{L}_{\text{enc}} = 
- \mathbb{E}_{X_t \sim \mathcal{D}_t}\!\left[\log D(f_{\text{enc}}^{T}(X'_t))\right].
\end{equation}
This adversarial process establishes a minimax game between the discriminator $D$ and the target encoder $f_{\text{enc}}^{T}$:
\begin{align}
\min_{f_{\text{enc}}^{T}} \max_{D} \;
\mathcal{L}_{\text{adv}}(f_{\text{enc}}^{T}, D)
&= \mathbb{E}_{X_s \sim \mathcal{D}_s}\!\left[\log D(f_{\text{enc}}(X'_s))\right] \nonumber \\
&\hspace{-3em} + \mathbb{E}_{X_t \sim \mathcal{D}_t}\!\left[\log (1 - D(f_{\text{enc}}^{T}(X'_t)))\right].
\end{align}
The adversarial loss $\mathcal{L}_{\text{adv}}$ aligns the feature distributions of the two domains.
Upon convergence, the target encoder learns to map real-world inputs to a latent feature distribution $P(Z_t)$ that approximates the simulated source distribution $P(Z_s)$. 
This yields a domain-invariant encoder $f_{\text{enc}}^{T}$， forming a bridge between simulation and reality and providing a stable foundation for subsequent fine-tuning on sparse UAV measurements.

\subsection{Real Data Finetuning}

After adversarial domain adaptation, the model obtains domain-invariant latent representations that effectively bridge the distributional gap between simulated and real-world data. 
Nevertheless, these aligned features still require further calibration to adapt to the fine-grained propagation characteristics of real environments. 
Therefore, the final stage of our framework, as illustrated in Fig.~\ref{fig:framework} (c), performs targeted finetuning using real sparse UAV measurements to achieve this domain-specific refinement.

In this stage, the domain-adapted encoder $f_{\text{enc}}^{T}(\cdot)$ is frozen to preserve the robust, physically grounded spatial representations learned from both simulation and adversarial alignment. 
Only the decoder $f_{\text{dec}}(\cdot)$ remains trainable, allowing the network to refine the mapping from latent features to RSRP predictions without overwriting learned priors.
Given a masked 3D grid $X'_{\text{ground}}=M_{\text{ground}}\odot X_t$ containing all ground and non-aerial measurements, 
the reconstruction is obtained as:
\begin{equation}
  \hspace{-1em}  \hat{X}_c^{(i)} , \hat{X}_c^{(j)} =
\big[f_{\text{dec}}^{(i)}\!\left(f_{\text{enc}}^{T}(X'^{(i)}_{\text{ground}})\right),
f_{\text{dec}}^{(j)}\!\left(f_{\text{enc}}^{T}(X'^{(j)}_{\text{ground}})\right)\big].
\end{equation}

The decoder is updated by minimizing the MSE loss between the reconstructed and observed RSRP values at the sparse aerial measurement locations

\begin{align}
\mathcal{L}_{\text{finetune}}
&= \frac{1}{|\mathcal{D}_a^{(i)}|}
\sum_{(\mathbf{x}_{a,m},p_{a,m})\in\mathcal{D}_a^{(i)}}
\big(\hat{X}_c^{(i)}(\mathbf{x}_{a,m}) - p_{a,m}\big)^2 \nonumber\\
&\hspace{-2em}\quad + \frac{1}{|\mathcal{D}_a^{(j)}|}
\sum_{(\mathbf{x}_{a,m},p_{a,m})\in\mathcal{D}_a^{(j)}}
\big(\hat{X}_c^{(j)}(\mathbf{x}_{a,m}) - p_{a,m}\big)^2.
\end{align}
By restricting the adaptation to the decoder, this stage achieves a principled balance between two objectives:
(i) preserving the propagation-consistent latent representations encoded in $f_{\text{enc}}^{T}$, 
and (ii) calibrating the output distribution to match real-world signal variations. 
This decoder-only finetuning effectively bridges the final gap between general knowledge and environment-specific adaptation, 
enabling accurate radio map reconstruction for each cell ID even under extremely sparse UAV measurements.

\section{Experiments}
\begin{figure}
    \centering
    \includegraphics[width=0.65\linewidth]{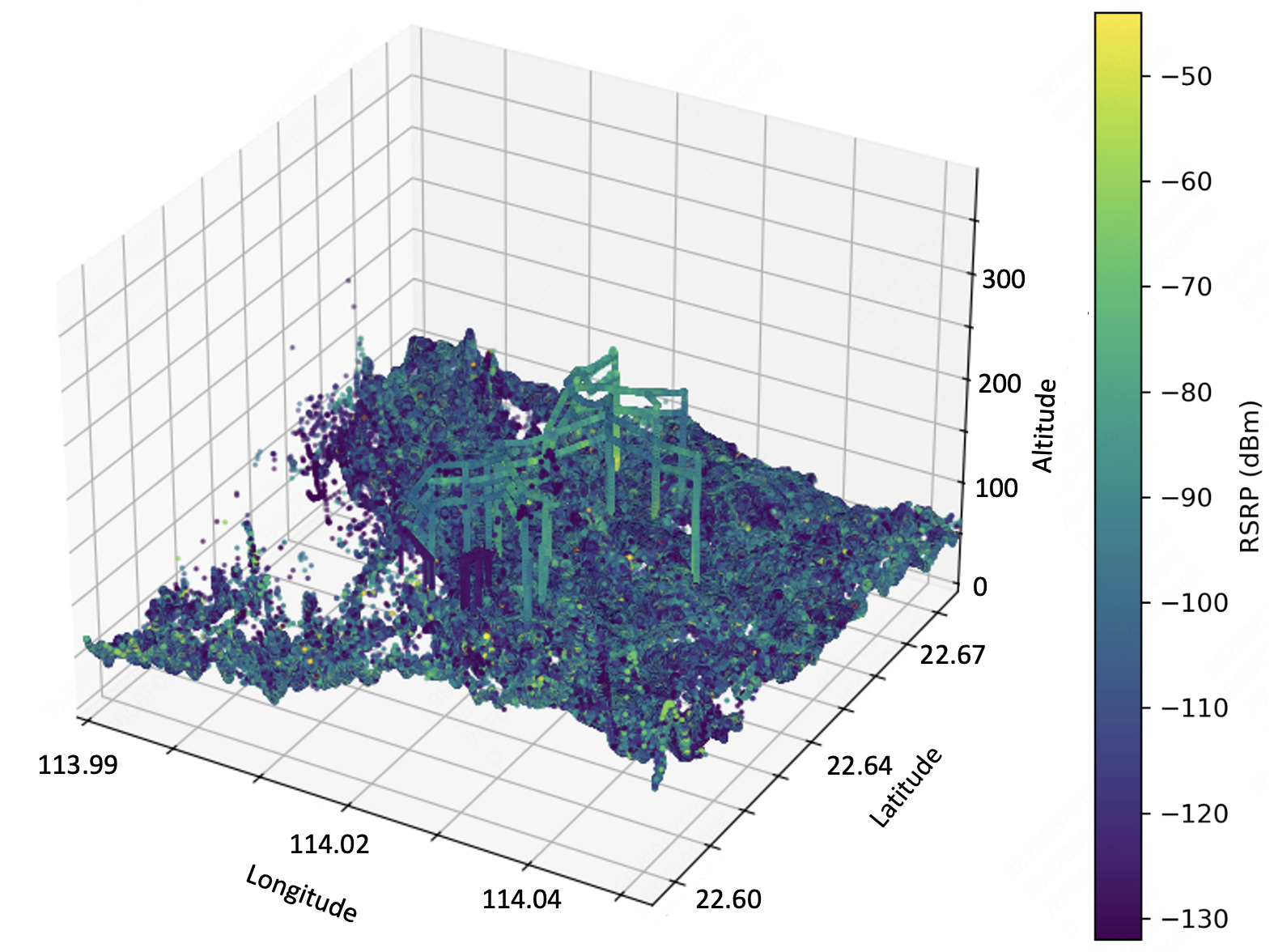}
    \caption{Visualization of a subset of the Meituan UAV collected aerial RSRP data along with ground crowdsourced RSRP measurements in an urban area.}
    \label{fig:realdata}
\end{figure}

\renewcommand{\baselinestretch}{1}\normalsize
This section presents a comprehensive evaluation of our proposed framework from three aspects (1) validating the simulation-based pretraining on synthetic data; (2) benchmarking our model against baselines on real UAV operational data (Fig.~\ref{fig:realdata}); and (3) an ablation study to isolate the impact of each core component. Performance is measured by root mean square error (RMSE) in decibles (dB).
\subsection{Data Augmentation via Raytracing}
At stage-1, we construct realistic $1000\times1000\,\mathrm{m}^2$ urban scenarios in the Sionna ray-tracing engine by importing building footprints and terrain data (OSM, DEM). In each scenario, 10 base stations are deployed, and the RSRP is computed on a dense $100\times100\times20$ 3D grid, yielding complete ground-to-air radio maps up to $200\,\mathrm{m}$ altitude. The resulting synthetic data are then used to pretrain our U\mbox{-}Net via a self-supervised masked reconstruction task, where the network learns to recover the full radio map from partially observed inputs; we set $(p_{\text{near}},p_{\text{mid}},p_{\text{far}})=(0.8,0.2,0.1)$ to balance realism and reconstruction difficulty. The ray-tracing configuration is summarized in Table~\ref{tab:rt_config}.

\begin{table}[h]
\centering
\caption{Raytracing configuration used to generate synthetic radio maps.}
\label{tab:rt_config}
\small
\begin{tabular}{c c}
\hline
\textbf{Parameter} & \textbf{Value} \\
\hline
Max path depth & 2 \\
Diffraction & \texttt{False} \\
Refraction & \texttt{False} \\
Grid cell size & $(10\,\mathrm{m},\,10\,\mathrm{m})$ \\
Samples per transmitter & $10^{9}$ \\
\hline
\end{tabular}
\end{table}

\begin{figure}[h!]
    \centering
    \includegraphics[width=1\linewidth]{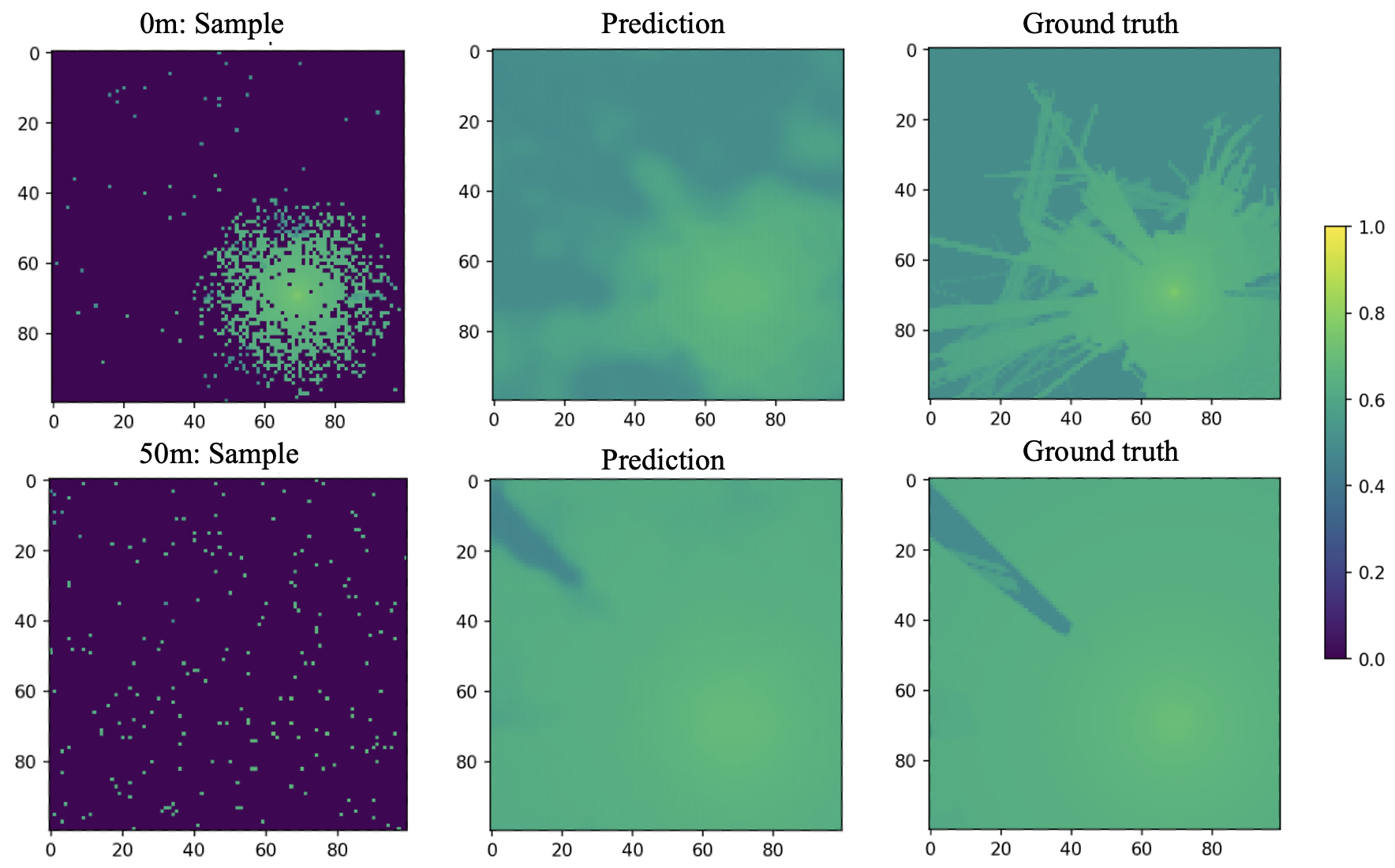}
    \caption{Masked U-Net reconstruction on simulated radio maps at 0\,m (top) and 50\,m (bottom). Columns show masked samples, predictions, and simulation ground truth.}
    \label{fig:sim_mask_results}
\end{figure}

The effectiveness of this pretraining is demonstrated in Fig.~\ref{fig:sim_mask_results}. The qualitative results show that the network successfully recovers fine-grained propagation patterns and restores large missing regions with physically consistent structures, even from sparse initial samples. This strong reconstruction performance on purely simulated data confirms that the pretraining process effectively instills a generalizable understanding of radio propagation, providing a solid initialization for the subsequent adaptation to real world.

\subsection{Performance Comparison}
To validate the effectiveness on real-world operational data of our proposed framework, we conduct a comprehensive quantitative comparison against several established baseline methods. The evaluation is performed on historical operational data collected from Meituan's UAV logistics platform.
Table~\ref{tab:compare} reports the RMSE statistics across Kriging \cite{kriging}, Gaussian Processes (GPs) \cite{GPs}, Autoencoder \cite{autoencoder}, 
and our proposed framework. 

Classical methods (Kriging and GPs) exhibit relatively high errors, 
especially in the worst-case routes, reflecting their limited ability to generalize to complex aerial propagation patterns. 
The Autoencoder baseline achieves moderate improvements but still struggles when training data is sparse. 
In contrast, our proposed framework substantially reduces RMSE across best, mean, and worst cases. 
This demonstrates that combining simulation-driven pretraining with adversarial domain adaptation provides robust representations, enabling reliable real world aerial radio map recovery.
\begin{table}[h]
\centering
\caption{Performance comparison on Shenzhen UAV test routes.}
\label{tab:compare}
\begin{tabular}{cccc}
\hline
\textbf{Method} & \textbf{Best} & \textbf{Mean} & \textbf{Worst} \\
\hline
Kriging \cite{kriging}         & 9.5   & 12.1  & 18.3 \\
GPs \cite{GPs}            & 7.6   & 11.5  & 19.7 \\
Autoencoder \cite{autoencoder}     & 7.8   & 10.2  & 16.9 \\
Proposed  & \textbf{3.4} & \textbf{5.3} & \textbf{12.7} \\
\hline
\end{tabular}
\end{table}

\subsection{Ablation Studies}
We conduct ablation studies to quantify the contribution of each component in our framework. 
The switch-style results on Shenzhen UAV test routes are summarized in Table~\ref{tab:perf_ablation_switch}. 
{
\begin{itemize}
    \item \textbf{w/o Simulation Pretraining:} Removing the simulation pretraining stage leads to a sharp increase in error 
    (mean RMSE rises from 5.3 to 9.5 dB). This degradation occurs because the model loses the propagation priors and structural awareness learned from large-scale synthetic radio maps. 
    Without these priors, the encoder cannot effectively capture the spatial correlation between signal strength and urban geometry, leading to poor generalization on real UAV data.

    \item \textbf{w/o Adversarial Domain Adaptation:} Disabling ADDA results in higher error 
    (mean RMSE increases from 5.3 to 5.8 dB). The absence of domain alignment causes the latent features extracted from real-world measurements to deviate from those learned on simulation data. 
    Consequently, the decoder receives inconsistent representations and fails to transfer the learned physical priors to the real domain, widening the sim-to-real gap.

    \item \textbf{w/o Real Data Finetuning:} Without finetuning on real UAV measurements, the performance degrades noticeably 
    (mean RMSE grows to 6.6 dB). Although adversarial adaptation provides domain-invariant features, the model remains uncalibrated to real-world signal dynamics like material-dependent reflections.

    \item \textbf{w/o Dual-cell Input:} Restricting the input to a single cell increases error (mean RMSE: 6.3 dB), as the model can no longer leverage the complementary signals from adjacent cells in overlapping coverage regions. This prevents the exploitation of inter-cell spatial consistency, which is vital for accurately inferring occluded areas.
\end{itemize}}
The full model, integrating all components, achieves the lowest errors 
(best/mean/worst RMSE = 3.4/5.3/12.7 dB), confirming the efficacy of our design.

\begin{table}[h]
\centering
\caption{Ablation study on Shenzhen UAV test routes.}
\label{tab:perf_ablation_switch}
\begin{tabular}{cccc|ccc}
\toprule
\textbf{Pretrain} & \textbf{ADDA} & \textbf{Finetuning} & \textbf{Dual-cell} & \textbf{Best} & \textbf{Mean} & \textbf{Worst} \\
\midrule
$\checkmark$ & $\checkmark$ & $\checkmark$ & $\checkmark$ & \textbf{3.4} & \textbf{5.3} & \textbf{12.7} \\
$\checkmark$ & $\times$     & $\checkmark$ & $\checkmark$ & 3.7 & 5.8 & 12.9 \\
$\checkmark$ & $\checkmark$ & $\times$     & $\checkmark$ & 4.5 & 6.6 & 13.5 \\
$\checkmark$ & $\checkmark$ & $\checkmark$ & $\times$     & 3.9 & 6.3 & 13.2 \\
$\times$     & $\times$     & $\checkmark$ & $\checkmark$ & 7.5 & 9.5 & 16.5 \\
$\checkmark$ & $\times$     & $\times$     & $\checkmark$ & 12.6 & 17.1 & 20.8 \\
\bottomrule
\end{tabular}
\end{table}

\begin{figure}[ht]
    \centering
    \includegraphics[width=1\linewidth]{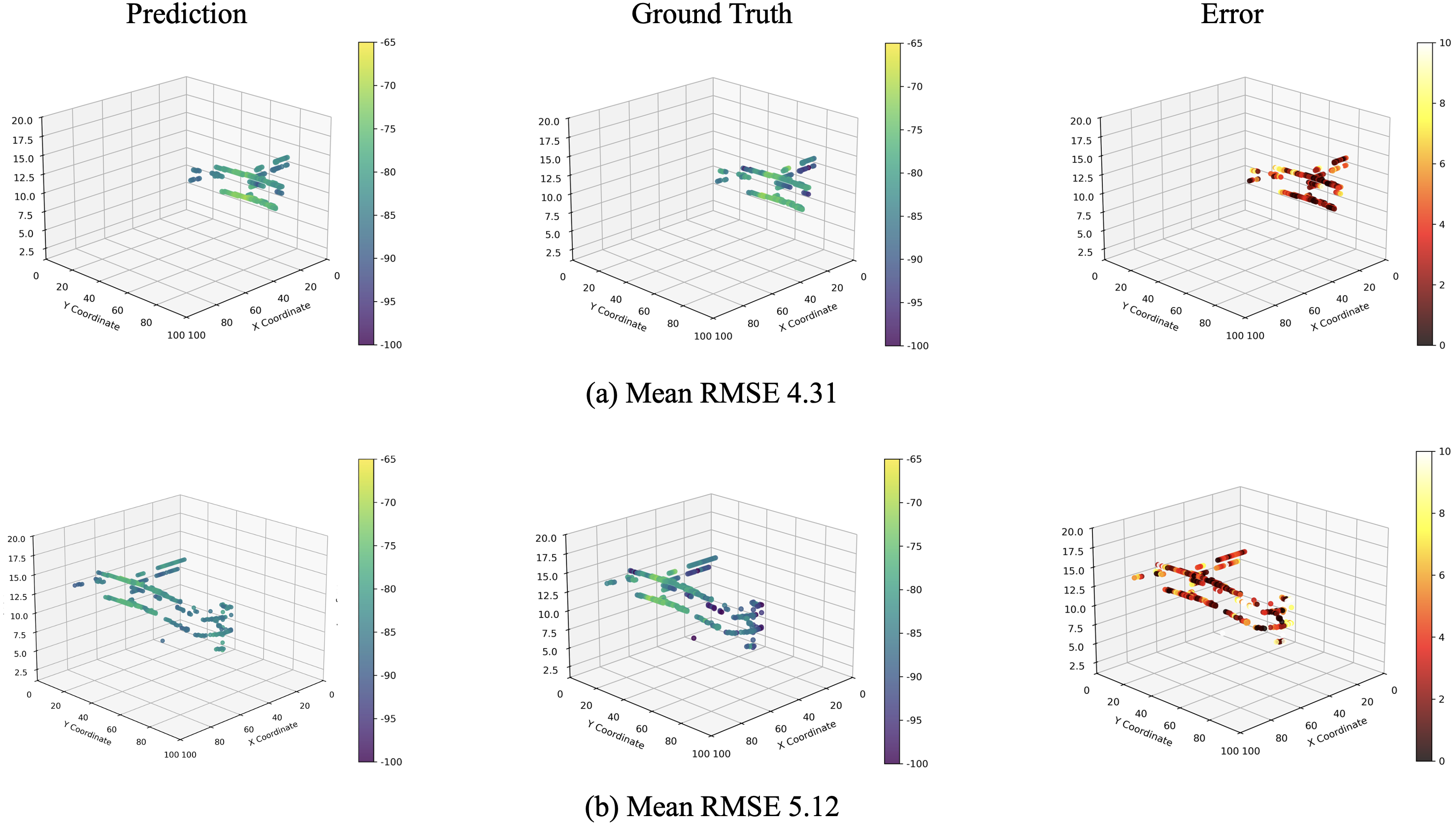}
    \caption{Predicted and ground truth RSRP profiles along real UAV test trajectory.}
    \label{fig:shenzhen_curves}
\end{figure}

\section{Conclusions}
We have demonstrated an effective approach for predicting route-level UAV connectivity by transferring knowledge from abundant ground measurements to the aerial domain. Our method's key innovation lies in its three-stage design: learning propagation priors from simulation, aligning feature distributions adversarially, and calibrating with minimal real UAV data. This framework achieved a 51.9\% improvement in accuracy on real operational data, offering a practical and scalable solution for reliable low-altitude communication planning.

{
\small                       
\renewcommand{\baselinestretch}{0.83}\normalsize
\bibliographystyle{IEEEtran}
\bibliography{reference}

@ARTICLE{matrixInterpolation,
  author={Sun, Hao and Chen, Junting},
  journal={IEEE Transactions on Signal Processing}, 
  title={{Propagation Map Reconstruction via Interpolation Assisted Matrix Completion}}, 
  year={2022},
  volume={70},
  number={},
  pages={6154-6169},
  doi={10.1109/TSP.2022.3230332}}

@INPROCEEDINGS{sip2net,
  author={Lu, Wenlihan and Lu, Ziyi and Yan, Jia and Gao, Shijian},
  booktitle={ICASSP 2025 - 2025 IEEE International Conference on Acoustics, Speech and Signal Processing (ICASSP)}, 
  title={{SIP2Net: Situational-Aware Indoor Pathloss-Map Prediction Network for Radio Map Generation}}, 
  year={2025},
  volume={},
  number={},
  doi={10.1109/ICASSP49660.2025.10890319}}

@ARTICLE{radiomap,
  author={Romero, Daniel and Kim, Seung-Jun},
  journal={IEEE Signal Processing Magazine}, 
  title={{Radio Map Estimation: A Data-driven Approach to Spectrum Cartography}}, 
  year={2022},
  volume={39},
  number={6},
  pages={53-72},
  doi={10.1109/MSP.2022.3200175}}

@misc{sionna,
      title={{Sionna: An Open-Source Library for Next-Generation Physical Layer Research}}, 
      author={Jakob Hoydis and Sebastian Cammerer and Fayçal Ait Aoudia and Avinash Vem and others},
      year={2023},
      eprint={2203.11854},
      archivePrefix={arXiv},
      primaryClass={cs.IT},
      url={https://arxiv.org/abs/2203.11854}, 
}

@INPROCEEDINGS{adda,
  author={Tzeng, Eric and Hoffman, Judy and Saenko, Kate and Darrell, Trevor},
  booktitle={2017 IEEE Conference on Computer Vision and Pattern Recognition (CVPR)}, 
  title={{Adversarial Discriminative Domain Adaptation}}, 
  year={2017},
  volume={},
  number={},
  pages={2962-2971},
  doi={10.1109/CVPR.2017.316}}

@ARTICLE{GPs,
  author={Zhen, Pan and Zhang, Bangning and Xu, Yi-Qun and Chen, Zhibo and others},
  journal={IEEE Wireless Communications Letters}, 
  title={{Radio Environment Map Construction Based on Gaussian Process With Positional Uncertainty}}, 
  year={2022},
  volume={11},
  number={8},
  pages={1639-1643},
  doi={10.1109/LWC.2022.3170147}}

@ARTICLE{autoencoder,
  author={Teganya, Yves and Romero, Daniel},
  journal={IEEE Transactions on Wireless Communications}, 
  title={{Deep Completion Autoencoders for Radio Map Estimation}}, 
  year={2022},
  volume={21},
  number={3},
  pages={1710-1724},
  doi={10.1109/TWC.2021.3106154}}

@misc{cbam,
      title={CBAM: Convolutional Block Attention Module}, 
      author={Sanghyun Woo and Jongchan Park and Joon-Young Lee and In So Kweon},
      year={2018},
      eprint={1807.06521},
      archivePrefix={arXiv},
      primaryClass={cs.CV},
      url={https://arxiv.org/abs/1807.06521}, 
}

@ARTICLE{radioUnet,
  author={Levie, Ron and Yapar, Cagkan and Kutyniok, Gitta and Caire, Giuseppe},
  journal={IEEE Transactions on Wireless Communications}, 
  title={{RadioUNet: Fast Radio Map Estimation With Convolutional Neural Networks}}, 
  year={2021},
  volume={20},
  number={6},
  pages={4001-4015},
  doi={10.1109/TWC.2021.3054977}}

@ARTICLE{spectrumsurveying,
  author={Shrestha, Raju and Romero, Daniel and Chepuri, Sundeep Prabhakar},
  journal={IEEE Transactions on Wireless Communications}, 
  title={{Spectrum Surveying: Active Radio Map Estimation With Autonomous UAVs}}, 
  year={2023},
  volume={22},
  number={1},
  pages={627-641},
  doi={10.1109/TWC.2022.3197087}}

@ARTICLE{OSM,
  author={Vargas-Munoz, John E. and Srivastava, Shivangi and Tuia, Devis and Falcão, Alexandre X.},
  journal={IEEE Geoscience and Remote Sensing Magazine}, 
  title={{OpenStreetMap: Challenges and Opportunities in Machine Learning and Remote Sensing}}, 
  year={2021},
  volume={9},
  number={1},
  pages={184-199},
  doi={10.1109/MGRS.2020.2994107}}

@INPROCEEDINGS{kriging,
  author={Gao, Ying and Fujii, Takeo},
  booktitle={2022 IEEE 96th Vehicular Technology Conference (VTC2022-Fall)}, 
  title={{Kriging-based Trust Nodes Aided REM Construction under Threatening Environment}}, 
  year={2022},
  volume={},
  number={},
  doi={10.1109/VTC2022-Fall57202.2022.10012983}}

@ARTICLE{erergency,
  author={Li, Bowen and Chen, Junting},
  journal={IEEE Transactions on Wireless Communications}, 
  title={{Radio Map-Assisted Approach for Interference-Aware Predictive UAV Communications}}, 
  year={2024},
  volume={23},
  number={11},
  pages={16725-16741},
  doi={10.1109/TWC.2024.3446240}}

@misc{monitoring,
  title        = {{Low-Altitude Wireless Networks: A Survey}},
  author       = {Jun Wu and Yaoqi Yang and Weijie Yuan and Wenchao Liu and others},
  year         = {2025},
  eprint       = {2509.11607},
  archivePrefix= {arXiv},
  primaryClass = {eess.SP},
  url          = {https://arxiv.org/abs/2509.11607},
}

@misc{activeradiomap,
      title={{Bayesian-Driven Graph Reasoning for Active Radio Map Construction}}, 
      author={Wenlihan Lu and Shijian Gao and Miaowen Wen and Yuxuan Liang and others},
      year={2025},
      eprint={2508.09142},
      archivePrefix={arXiv},
      primaryClass={eess.SP},
      url={https://arxiv.org/abs/2508.09142}, 
}

@ARTICLE{machineradio,
  author={Zhang, Songyang and Choi, Brian and Ouyang, Feng and Ding, Zhi},
  journal={IEEE Communications Magazine}, 
  title={{Physics-Inspired Machine Learning for Radiomap Estimation: Integration of Radio Propagation Models and Artificial Intelligence}}, 
  year={2024},
  volume={62},
  number={8},
  pages={155-161},
  doi={10.1109/MCOM.001.2300782}}

@ARTICLE{cellswitch,
  author={Bernabè, Matteo and López-Pérez, David and Piovesan, Nicola and Geraci, Giovanni and others},
  journal={IEEE Open Journal of the Communications Society}, 
  title={{Massive MIMO for Aerial Highways: Enhancing Cell Selection via SSB Beams Optimization}}, 
  year={2024},
  volume={5},
  number={},
  pages={3975-3996},
  doi={10.1109/OJCOMS.2024.3418339}}

@ARTICLE{rmgen,
  author={Luo, Xuanhao and Li, Zhizhen and Peng, Zhiyuan and Chen, Mingzhe and others},
  journal={IEEE Transactions on Cognitive Communications and Networking}, 
  title={{Denoising Diffusion Probabilistic Model for Radio Map Estimation in Generative Wireless Networks}}, 
  year={2025},
  volume={11},
  number={2},
  pages={751-763},
  doi={10.1109/TCCN.2025.3529879}}

@ARTICLE{temporal,
  author={Zhang, Shuowen and Zhang, Rui},
  journal={IEEE Transactions on Wireless Communications}, 
  title={{Radio Map-Based 3D Path Planning for Cellular-Connected UAV}}, 
  year={2021},
  volume={20},
  number={3},
  pages={1975-1989},
  doi={10.1109/TWC.2020.3037916}}
}
\end{document}